\documentclass[journal]{IEEEtran}

\IEEEoverridecommandlockouts                            
\usepackage{centernot}
\usepackage[noadjust]{cite}
\usepackage{longtable}
\usepackage{amsmath}
\usepackage{graphicx, amssymb}
\usepackage[dvips]{epsfig}
\usepackage{color}
\usepackage[dvipsnames]{xcolor}
\usepackage{comment}
\usepackage{balance}
\usepackage{tikz}
\usepackage{enumitem}

\def\tt{\tilde \theta}

\def\L2{{\cal L}_2}

\newlength{\defbaselineskip}
\newcommand{\setlinespacing}[1]%
           {\setlength{\baselineskip}{#1 \defbaselineskip}}

\newcommand{\beq}{\begin{equation}}
\newcommand{\eeq}{\end{equation}}

\makeatletter
\newcommand{\xleftrightarrow}[2][]{\ext@arrow 3359\leftrightarrowfill@{#1}{#2}}
\makeatother

\newcommand{\xdasharrow}[2][-->]{
\tikz[baseline=-\the\dimexpr\fontdimen22\textfont2\relax]{
\node[anchor=south,font=\scriptsize, inner ysep=1.5pt,outer xsep=8pt](x){#2};
\draw[shorten <=3.4pt,shorten >=3.4pt,dashed,#1](x.south west)--(x.south east);
}
}

\def\BibTeX{{\rm B\kern-.05em{\sc i\kern-.025em b}\kern-.08em
    T\kern-.1667em\lower.7ex\hbox{E}\kern-.125emX}}

\smallskip

 \newtheorem{thm}{Theorem}

\newtheorem{ass}{Assumption}

\newtheorem{rmk}{Remark}

\allowdisplaybreaks

\title{ A Nonlinear Controller for Parallel DC-DC Converters  with   ZIP Load and Constrained Output Voltage}

\author{ Somayyeh Bahrami$^{\dagger}$
\thanks{$^{\dagger}$ The author is with the Department of Electrical Engineering, Razi University, Kermanshah, Iran, \texttt{\small s.bahrami@razi.ac.ir}}%
}

\begin{document}

\maketitle

\thispagestyle{plain}
\pagestyle{plain}

\begin{abstract}
In this paper,   an adaptive nonlinear controller is designed for a parallel DC-DC converter system that feeds an unknown  ZIP load, characterized by constant impedance (Z), constant current (I), and constant power (P),  at the DC bus. The proposed controller ensures simultaneous voltage adjustment and power sharing in the large signal sense despite uncertainties in ZIP loads, DC input voltages, and other electrical parameters. To keep the output voltage within a desired range, we utilize a barrier function that is invertible, smoothly continuous, and strictly increasing. Its limits at infinity represent the upper and lower bounds for the output voltage. We apply the invertible transformation of the barrier function to the output voltage and then design the controller using the adaptive backstepping method. Using this barrier-function-based
 adaptive backstepping controller,  uncertain parameters are identified on-line,  
and the  voltage adjustment and power sharing objectives are established. Moreover, voltage constraint is not violated event in the presence of  sudden and unknown large  variations of load. 
The efficiency of the proposed nonlinear controller is evaluated through simulations of a parallel DC-DC converter system using the MATLAB/Simscape Electrical environment.
 
\end{abstract}

\section{Introduction}
Because of the limited output current of a single DC-DC converter, the parallel interconnection of multiple converters has received significant attention over the past few decades. As parallel converters are widely employed in low-voltage/high-current applications with high performance, the development of appropriate control strategies to guarantee stability and reliability becomes increasingly crucial.
In such parallel converter systems, the main control objective is to achieve voltage adjustment and current sharing, simultaneously. Several control methods have been proposed in the literature to accomplish this objective, which  can be classified into droop (e.g. \cite{8616804, Anand, 988666}) and non-droop categories (e.g. \cite{704129,6851919,4195635, 6727450, 4418525, 6109354, 5457984, 4745797, 5613924, 8269340, TREGOUET201959, 8960535, 8269340, 8064725  } ).In droop approaches, a virtual resistance is added to the output characteristic of each converter. However, since the output voltage droops by increasing the load current, droop methods often result in poor voltage adjustment \cite{988666}. Active current-sharing schemes, which generally have an outer loop for voltage control and an inner loop for current control, are widely employed for parallel DC-DC converter systems \cite{704129,6851919,4195635, 6727450, 4418525, 6109354, 5457984, 4745797, 5613924, 8269340}.  To decouple the dynamics of these control loops, a frequency separation argument is necessary, which inevitably limits achievable performance \cite{TREGOUET201959}. In \cite{TREGOUET201959}, a geometrical decomposition-based method has been presented, where no frequency consideration  is needed to decouple the dynamics of voltage regulation and current sharing. In \cite{8960535}, based on a Hamiltonian framework, it has been shown that separation between current distribution and voltage regulation loops is related to the Casimir functions, and then a robust controller has been proposed. The controller proposed in \cite{8269340} ensures power sharing and voltage regulation while minimizing total losses in the presence of load uncertainty. In \cite{8064725}, a hybrid communication scheme comprising a centralized scheme for current sharing and a decentralized scheme for load voltage regulation has been proposed.
  
  All of the aforementioned works have considered Z or I loads. However, in recent decades, owning advancements in power electronic devices, a significant proportion of loads in DC microgrids consist of P-loads. These loads impose a negative incremental impedance on the DC microgrid system, which can greatly reduce the damping rate and even lead to instability of the entire system \cite{Singh2017}. In \cite{9201011},   the author considers a parallel DC-DC converter system feeding a ZIP load and proposes a control approach to ensure local asymptotic stability, albeit under a restrictive condition on the loads. Furthermore, several averaging-based control methods for DC microgrids with ZIP loads have been proposed in \cite{9611083, 10471262, 8836491, 9279289, DEPERSIS2018364}. These methods aim to achieve proportional power sharing and regulation of the weighted geometric mean of bus voltages, but they are effective only within a limited range of loads.

Indeed, maintaining voltage within prescribed bounds during transient operation in DC microgrids is crucial to prevent equipment damage. However, only a few articles in the literature address this issue \cite{9409144, 9991257, 10412654}. In \cite{9409144}, an optimization-based controller has been designed using control barrier functions to achieve safe proportional current sharing and weighted average bus voltage regulation in DC microgrids with unknown Z-load.  

In this paper, we consider a parallel DC-DC converter system feeding  a ZIP load. There are two challenges here. First, P-loads introduce high-order nonlinearity to the system, necessitating the use of efficient nonlinear controllers to ensure large-signal stability, especially in the presence of significant perturbations such as large unknown load variations. Second, since system parameters such as load values and electrical elements are always unknown, the controller should not rely on precise values of such parameters being available. To keep the output voltage within
a desired range, we employ a barrier function, which is invertible,
smoothly continuous and strictly increasing, and its limits at
infinity are the considered upper and lower bounds for the
output voltage.  
We apply the invertible transformation
of the barrier function on the output voltage and design the
local controller of each converter using the adaptive backstepping
method. 
The main
contributions of this article in comparison to  the
existing nonlinear voltage controllers in the literature  are as follows:
\begin{itemize}
\item Differently from the passivity-based nonlinear controllers in \cite{Soloperto2018},  \cite{24nahata2020} and \cite{9134402}, which guarantee only voltage regulation  inside a region of
attraction depending on the Z-load and P-load,  we ensure  voltage regulation and power sharing globally in a parallel DC-DC converter system feeding a ZIP load   without any dependency and restrictive conditions on the loads. 
\item  By integrating a barrier function into our Lyapunov-based controller design, we ensure constraint satisfaction during transient response, even in the presence of parameter uncertainties and unknown large load variations.
\item  To the best of our knowledge, this is the first time in the literature that a nonlinear controller is proposed for the voltage regulation and power sharing without measuring the DC input voltages of the converters. In our method, the DC input voltage of each converter is estimated using a proposed adaptation law. In fact, due to the use of fewer sensors, our control scheme is less expensive and more reliable.      \end{itemize}
The paper is organized as follows. In Section II, a dynamic model of the considered parallel DC-DC converter system is presented, and the main control problem is expressed. Section III is dedicated to developing the barrier-function-based adaptive backstepping controller. In Section IV, the performance of the proposed controller is illustrated via numerical simulation. Finally, Section V includes some concluding remarks.
 \emph{Notation:} $\mathbb{R}$ denotes the set of real numbers. The transpose of matrix $M$ is denoted  by $M^T$.  
   \section{problem formulation} 
 In this section, we describe  the dynamic model of the
considered system and then state  the main control
objectives.
\subsection{DC microgrid dynamics}
\begin{figure}
 \centering
   \includegraphics[width=8.5 cm]{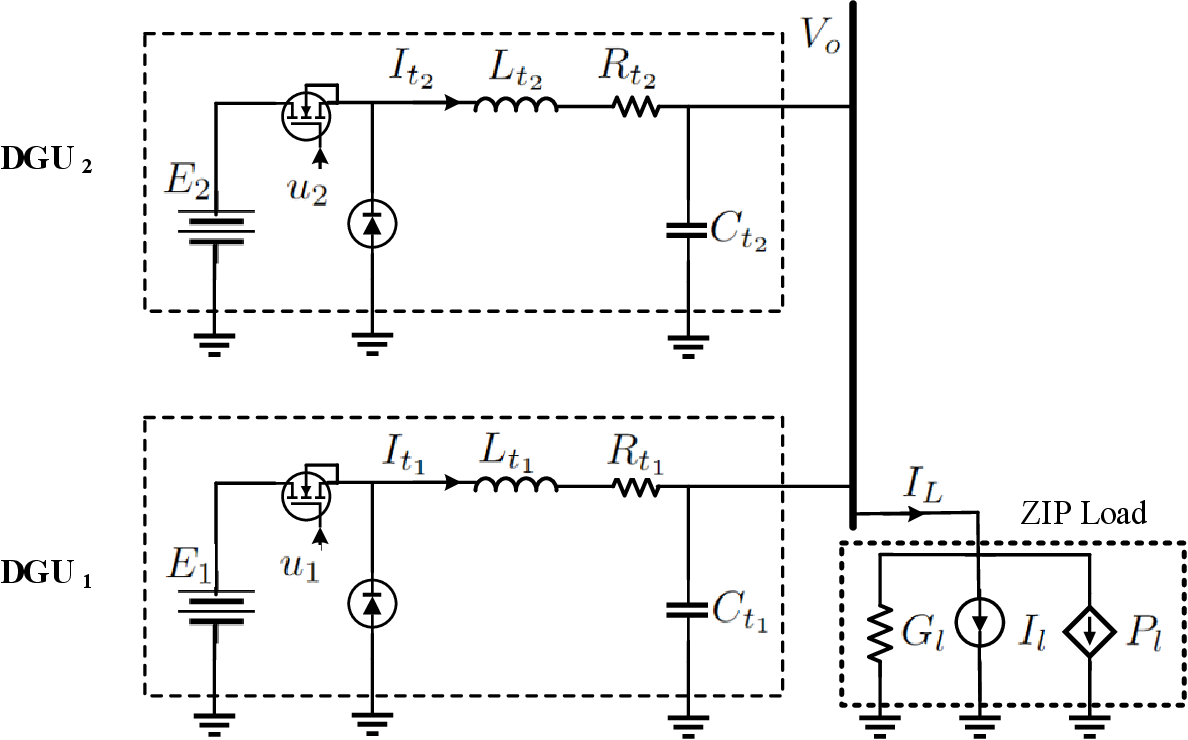} \vspace*{-0.2cm}
  \caption{ Electrical scheme of a  typical  DC microgrid  system} \label{Fig_2DG}
 \end{figure}

Consider an islanded DC microgrid  containing  $n$ parallel distributed generation units (DGUs), which feeds  a  ZIP load   at the DC bus. 
Fig.~\ref{Fig_2DG}  shows the electrical scheme of such a DC microgrid system with two DGUs where each DGU includes   a DC  voltage source,   a DC-DC converter, and a   resistive-inductive-capacitive filter. 
 Based on an average model,  the dynamic model of  each DGU  can be  written as
\begin{align}
\mathcal{C}_{t} \dot{V}_{o}&= I_{t}-G_{l }V_{o }-I_{l}-V_{o}^{-1}P_{l}  \label{dgi}\\
L_{t_i}\dot{I}_{t_i}&=-V_{o}-R_{t_i}I_{t_i}+E_iu_i\label{iti}
\end{align}
for $i=1,...,n$, where the applied symbols are described  in Table I,  $I_{t}=\Sigma_{i=1}^n I_{t_i}$, and $\mathcal{C}_{t}=\Sigma_{i=1}^n C_{t_i}$.

\begin{table}
\begin{center}
\caption{Description of the applied symbols}
\begin{tabular} {|p{1cm}|p{2.3cm}| p{1cm}|p{2.75cm}| }
\hline
Symbol & Description & Symbol & Description \\
\hline
 $V_{o} $ & Output voltage &$R_{t_i} $ & Filter resistance \\
 $I_{t_i} $ & Filter  current&$L_{t_i} $ & Filter inductance \\
 $u_i $ &Control input&$C_{t_i} $ & Filter capacitance\\ $I_{l } $ & Current of I-load  &$G_{l} $ &  Conductance of Z-load\\
$P_{l } $ &Power of P-load
 &$E_i $ &  DC Input voltage \\
\hline
\end{tabular}
\end{center}
\end{table}

\subsection{Control objective}\label{CO}
To consider some uncertainty sources,  we make the following assumption:
\begin{ass} \label{ass_load} We assume    that  the load parameters $P_{l}$, $G_{l}$, $I_{l}$,  the  DC input voltage $E_i$,  the internal resistance  $R_{t_i}$, the   inductance $L_{t_i}$ and capacitance $C_{t_i}$ for  $i=1,2,...,n$ are    all unknown.
\end{ass}
The  objective of this paper is to design the control law $u_i$  under Assumption 1    such that the following control objectives are achieved, simultaneously:\\
\emph{Voltage regulation:}  The output voltage $V_{o}$  
converges to the given setpoint $V_{o}^*>0$ asymptotically while satisfying the constraint  $v_{min}< V_{o} <v_{max}$ with $ v_{max}> v_{min}>0$ for all $t \geq 0$.\\
\emph{Current sharing:} The total current demand is proportionally
distributed amongst the converters at the steady state, that is, 
\begin{align}
\lim_ {t\rightarrow \infty} I_{t_i}=r_i I_{L}^*\label{ppiti}
\end{align}
for $i=1,...,n$ where $\Sigma_{i=1 }^{n}r_{i}=1$ with $0<r_i<1$. Also, $I_{L}^* $ is the steady-state current demand and in the case of ZIP load is equal to 
$G_{l }V_{o }^*+I_{l}+\dfrac{P_{l}}{V_{o}^{*}}$.
In fact, to improve the generation efficiency, it is
generally desired that the total current demand is shared among
the various paralleled   DGUs       proportionally to the generation capacity
of their corresponding energy sources. This desire can be expressed as (\ref{ppiti})  for $i=1,...,n$,  where $r_{i}$ relates to the generation capacity of
$\textmd{DGU}_{i}$.

\section{Control Design}\label{section_control}
In this section, we design  a     controller based on  the two-step adaptive backstepping method to achieve  the aforementioned control objectives.
\subsection{Barrier-function-based  transformation}
To deal with the voltage  constraint $v_{min}< V_{o} < v_{max}$,  we consider the invertible transformation of
barrier functions as follows 
  \begin{align}
      V_o=\mathcal{F}(\mathcal{V})\Longleftrightarrow \mathcal{V}=\mathcal{F}^{-1}(V_o)\label{b1}
  \end{align}
  where
$\mathcal{F}(.)$ and $\mathcal{F}^{-1}(.)$
are inverse functions of each other. Moreover, they are smoothly continuous
and strictly increasing in their respective arguments, and satisfy
\begin{align}
\begin{cases}
v_{min} < \mathcal{F}(\mathcal{V})<v_{max}\\
\lim_{\mathcal{V}\rightarrow \infty} \mathcal{F}(\mathcal{V})=v_{max}\\
\lim_{\mathcal{V}\rightarrow -\infty} \mathcal{F}(\mathcal{V})=v_{min}
\end{cases},
    \begin{cases}
 -\infty< \mathcal{F}^{-1}( V_o)<\infty\\
 \lim_{V_o\rightarrow v_{max}} \mathcal{F}^{-1}(V_o)=\infty\\
\lim_{V_o\rightarrow v_{min}} \mathcal{F}^{-1}(V_o)=-\infty
\end{cases} \label{b2}
\end{align}
\begin{rmk} \label{rem1}
Evidently, if initial voltage   satisfies $v_{min} <  V_{o}(0) < v_{max}$, then $\mathcal{V}(0)$
is bounded. Furthermore, if $\mathcal{V}(t)$
is stabilized to be bounded for all $t\geq 0$, then $ V_o=\mathcal{F}(\mathcal{V})$
 satisfies the voltage constraint $v_{min}<  V_{o}(t) < v_{max}$ for all
$t \geq 0$. 
\end{rmk}

\subsection{First step of backstepping design}\label{sec31}

As the design objective is to enforce $V_{o}$ 
 and $I_{t_i}$ to converge to their desired values $V_{o }^*$ 
 and $I_{t_i}^*$
asymptotically,  the errors $Z_1$,   $Z_2$ and $Z_{2i}$ are defined as follows:
\begin{align}
   & Z_1=\mathcal{V}-\mathcal{V}^*\label{itild}\\
   &Z_{2}=I_{t}-\xi \label{vtild}\\
  & Z_{2_i}=I_{t_i}-\hat I_{t_i}^*\label{z2}
\end{align}
where considering $\mathcal{F}(.)$ as a barrier function with the properties mentioned in Section III-A, $\mathcal{V}=\mathcal{F}^{-1}(V_{o})$ and $\mathcal{V}^*=\mathcal{F}^{-1}(V_{o}^*)$. Evidently, converging $\mathcal{V}$ to $\mathcal{V}^*$ implies that $V_o$ also converges to $V_o^*$.   $\xi$ is a virtual control input which is determined in the sequel. Also, based on control objective (\ref{ppiti}), we consider $\hat I_{t_i}^*=r_i \hat I_L^*$ where $\hat I_L^*$ is the estimation of the steady-state current demand $I_L^*$. 

We define the unknown vector 
   $\Theta \in\mathbb{R}^{ 3}$ and the regressor matrix  
  $\Psi (V_{o})\in\mathbb{R}^{1 \times 3}$ as
  \begin{align}
&\Theta =\begin{bmatrix}G_{l }&P_{l }&I_{l }   \end{bmatrix}^T\label{lmi1}\\
&\Psi (V_{o })=\begin{bmatrix}V_{o }&V_{o }^{-1}&1 \end{bmatrix}
\label{lmi}
\end{align}
to reformulate   the unknown  current demand $I_{L} =G_{l }V_{o }+I_{l}+V_{o}^{-1}P_{l}$   as $I_{L}=\Psi (V_{o })\Theta$. Using this, (\ref{dgi}) is rewritten as
\begin{align}
     \dot{ V}_{o }= \mathcal{C}_{t }^{-1}I_{t } -\mathcal{C}_{t }^{-1}\Psi (V_{o })\Theta \label{dgi1}
\end{align}
By virtue of (\ref{dgi1}), (\ref{itild}) and $\mathcal{V}=\mathcal{F}^{-1}(V_{o})$, the dynamics of the error $Z_1$ is obtained as
\begin{align}
    \mathcal{C}_{t}\dot Z_1= \mathcal{C}_{t} \dfrac{d \mathcal{F}^{-1}(V_o)}{dV_o}  \dot V_o= \dfrac{d \mathcal{F}^{-1}(V_o)}{dV_o} \Big(I_{t}-\Psi (V_{o })\Theta  \Big)\label{Z1}
\end{align}
 In light of  (\ref{vtild}), we replace $I_t$ in (\ref{Z1})  with $Z_2+\xi$ and rewrite  (\ref{Z1}) as
\begin{align}
    \mathcal{C}_{t}\dot Z_1=  \dfrac{d \mathcal{F}^{-1}(V_o)}{dV_o} \Big(Z_2+\xi-\Psi (V_{o })\Theta\Big)\label{z22}
\end{align}
To determine the virtual input $\xi$, let us consider the following Lyapunov candidate:
\begin{align}
    W_1  =& \frac{1}{2}\mathcal{C}_{t } Z_1^2 +\frac{1}{2} \gamma_{1 }^{-1}\tilde \Theta^T \tilde \Theta \label{W1}
\end{align}
where $\gamma_{1 }$ is  a positive scalar and $\tilde \Theta =\Theta -\hat \Theta $ with $\hat \Theta $  as the estimated value of $\Theta $.
Differentiating $W_1$ with  respect  to  time and then substituting for $\dot{Z}_{1}$  from (\ref{z22}) yield that
\begin{align}
    \dot W_1 & =  Z_{1}\dot{Z}_{1}+\gamma_{1}^{-1}\tilde \Theta^T \dot{\hat\Theta}\nonumber\\&= Z_{1} \dfrac{d \mathcal{F}^{-1}(V_o)}{dV_o} \Big(Z_2+\xi-\Psi (V_{o })\Theta\Big)-\gamma_{1 }^{-1}\tilde \Theta ^T \dot{\hat \Theta}
     \label{dW1}
\end{align}
Now,       we design $\xi$ as  
      \begin{align} 
  \xi&=-\kappa_1\Big(\dfrac{d \mathcal{F}^{-1}(V_o)}{d V_o}\Big)^{-1} Z_1 +\Psi(V_{o}) \hat \Theta\label{x2ref}
\end{align}
where $\kappa_1$ is a positive scalar. Replacing $\xi$ in (\ref{dW1}), we get \begin{align}
    \dot W_1  =&  -\kappa_1 Z_{1}^2 +\dfrac{d \mathcal{F}^{-1}(V_o)}{dV_o}Z_1Z_2 -\dfrac{d \mathcal{F}^{-1}(V_o)}{dV_o}\Psi (V_{o })Z_1\tilde \Theta \nonumber\\&-\gamma_{1 }^{-1}\tilde \Theta ^T \dot{\hat \Theta}
     \label{ddW1}
\end{align}
Now, we assume that $\hat \Theta $ is calculated  based on the following adaptation law:
 \begin{align}
    \dot {\hat \Theta} =- \gamma_{1 } \dfrac{d \mathcal{F}^{-1}(V_o)}{dV_o} \Psi (V_{o })^T Z_1\label{adtet}
\end{align}
Substituting (\ref{adtet}) in (\ref{ddW1}), yields
\begin{align}
    \dot W_1  =&  -\kappa_1 Z_{1}^2 +\dfrac{d \mathcal{F}^{-1}(V_o)}{dV_o}Z_1Z_2 
     \label{dddW1}
\end{align}
where the term $\dfrac{d \mathcal{F}^{-1}(V_o)}{dV_o}Z_1Z_2 $ is compensated  in the 
next step.
\subsection{Second step of backstepping design}
First, we extract the dynamics of the error $Z_2$  defined in (\ref{vtild}).  
Taking the time derivative of (\ref{vtild}), we have 
\begin{align}
    \dot{Z}_2=\dot I_t-  \dot \xi\label{2z} 
\end{align}
In light of (\ref{iti}),  we   easily get the dynamics of $I_t=\Sigma_{i=1 }^{n}I_{t_{i }}$ as 
 \begin{align}
 &\dot{I}_{t }=- \Sigma_{i=1 }^{n}L_{t_{i }}^{-1}V_{o}-\Sigma_{i=1 }^{n}\dfrac{R_{t_{i}}}{L_{t_{i}}}I_{t_{i}}+\Sigma_{i=1 }^{n}\dfrac{E_{i}}{L_{t_{i }}}u_{i},\label{2ppiti}
 \end{align}
      Also, differentiating $\xi$ in (\ref{x2ref}) with respect to time, we have   \begin{align}
    \dot \xi &=\kappa_1\Big(\dfrac{d \mathcal{F}^{-1}(V_o)}{d V_o}\Big)^{-2} \dfrac{d^2 \mathcal{F}^{-1}(V_o)}{d V_o^2}\dot V_o Z_1 -\kappa_1 \dot V_o   \nonumber\\&+ \Psi (V_{o }) \dot{\hat \Theta} + \frac{d}{dt}\Psi (V_{o }) \hat \Theta \label{pvir} 
\end{align}
 Based on (\ref{lmi1}) and (\ref{lmi}),     $\frac{d}{dt} \Psi (V_{o })=\begin{bmatrix} \dot V_{o }&-V_{o }^{-2}\dot V_{o }&0 \end{bmatrix}$ and thus   \begin{align}
    \frac{d}{dt} \Psi (V_{o })\hat \Theta= \hat G_{l }\dot V_{o }-V_{o }^{-2} \hat P_{l }\dot V_{o }\label{ph}
\end{align}   
Substituting (\ref{ph})    in (\ref{pvir}) yields  that  
 \begin{align}   \dot \xi= & \Phi \dot{ V}_{o } +\Psi (V_{o }) \dot{\hat \Theta} \label{ppvir}
\end{align}
where \begin{align} \Phi =\kappa_1\Big(\dfrac{d \mathcal{F}^{-1}(V_o)}{d V_o}\Big)^{-2} \dfrac{d^2 \mathcal{F}^{-1}(V_o)}{d V_o^2} Z_1 -\kappa_1   + \hat G_{l }-V_{o }^{-2} \hat P_{l }\label{fii}\end{align}
Replacing $ \dot{ V}_{o }$ in (\ref{ppvir})  from (\ref{dgi1}) yields  \begin{align}   \dot \xi= & \Phi I_{t }\mathcal{C}_{t }^{-1} -\Phi\Psi (V_{o })\Theta_{c }+\Psi (V_{o }) \dot{\hat \Theta}\label{uvdf}
\end{align}
 where  $\Theta_{c }:=\Theta \mathcal{C}_{t }^{-1}$. 
   Finally, substituting (\ref{2ppiti}) and (\ref{uvdf})  into (\ref{2z}), we  derive   the  dynamics of   $Z_2$     as
\begin{align}
 \dot{Z}_{2}=&-\Sigma_{i=1 }^{n}L_{t_{i }} ^{-1}V_{o}-\Sigma_{i=1 }^{n}\lambda_iI_{t_{i}}+\Sigma_{i=1 }^{n}\mu_iu_{i}-\Phi I_{t }\mathcal{C}_{t }^{-1}\nonumber\\&+\Phi\Psi (V_{o })\Theta_{c }-\Psi (V_{o }) \dot{\hat \Theta},\label{52ppiti}
 \end{align}
 where $\lambda_i:=\dfrac{R_{t_{i}}}{L_{t_{i}}}$ and  $\mu_i:=\dfrac{E_i}{L_{t_{i}}}$. Defining $u=\Sigma_{i=1 }^{n}\hat \mu_iu_{i}$ and $\tilde \mu_i=\mu_i-\hat \mu_i$, we can rewrite (\ref{52ppiti}) as
\begin{align}
 \dot{Z}_{2}=&-\Sigma_{i=1 }^{n}L_{t_{i }} ^{-1}V_{o}-\Sigma_{i=1 }^{n}\lambda_iI_{t_{i}}+\Sigma_{i=1 }^{n}\tilde\mu_iu_{i}+u\nonumber\\&-\Phi I_{t }\mathcal{C}_{t }^{-1}+\Phi\Psi (V_{o })\Theta_{c }-\Psi (V_{o }) \dot{\hat \Theta}\label{42ppiti}
 \end{align}
 Also, in light of (\ref{iti}) and (\ref{z2}) and considering  $\hat I_{t_i}^*=r_i\hat I_L^*=r_i\Psi(V_{o }^*)  \hat \Theta$,  we obtain 
 \begin{align}
 \dot{Z}_{2i}=&-  L_{t_{i }}^{-1} V_{o}- \lambda_iI_{t_{i}}+ \tilde \mu_iu_{i}+ \hat\mu_iu_{i} -r_i\Psi(V_{o }^*)  \dot{\hat \Theta} \label{32ppiti1}
 \end{align}
Based on  (\ref{52ppiti}) and (\ref{32ppiti1}),   the  control laws of $u $ and $u_i$ for  $i=1,...,n-1$  are proposed  as:
\begin{align}
   u=& -\dfrac{d \mathcal{F}^{-1}(V_o)}{dV_o}Z_1-\kappa_2 Z_2+\Sigma_{i=1 }^{n}\hat L_{t_{i }} ^{-1}V_{o}+\Sigma_{i=1 }^{n}\hat \lambda_iI_{t_{i}}\nonumber\\&+\Phi I_{t }\hat {\mathcal{C}}_{t }^{-1}-\Phi\Psi (V_{o })\hat\Theta_{c }+\Psi (V_{o }) \dot{\hat \Theta}\label{nu_final}\\u_i=&\hat \mu_i^{-1}\Big(-\kappa_{2i}Z_{2i}+\hat  L_{t_{i }}^{-1} V_{o}+\hat \lambda_iI_{t_{i}}+r_i\Psi(V_{o }^*) \dot{\hat \Theta} \Big)\label{uii} 
\end{align}
where $\kappa_2$ and $\kappa_{2i}$ are positive  scalars. Now, based on $u=\Sigma_{i=1 }^{n}\hat \mu_iu_{i}$, (\ref{nu_final}) and (\ref{uii}), we determine $u_n$ as
\begin{align}
   u_n=& \hat \mu_n^{-1}\Big(-\dfrac{d \mathcal{F}^{-1}(V_o)}{dV_o}Z_1-\kappa_2 Z_2+\Sigma_{i=1 }^{n-1}\kappa_{2i}Z_{2i}+\hat L_{t_{n }} ^{-1}V_{o}\nonumber\\&\quad\quad\quad+\hat \lambda_nI_{t_{n}}+\Phi I_{t }\hat {\mathcal{C}}_{t }^{-1}-\Phi\Psi (V_{o })\hat\Theta_{c }\nonumber\\&\quad\quad\quad-\Sigma_{i=1 }^{n-1}r_i  \Psi (V_{o }^*) \dot{\hat \Theta}+\Psi (V_{o }) \dot{\hat \Theta}\Big)\label{un}
\end{align}
Also,  the estimated values $ \hat \Theta_{c}$,  $ {\hat{\mathcal{C}}}_{t}^{-1}$, ${\hat{L}}_{t_i}^{-1}$, $\hat \lambda_i$ and $\hat \mu_i$  for $i=1,...,n$ are calculated by the following adaptation laws: 
\begin{align}
    &\dot{\hat \Theta}_{c}=\gamma_{2 }  \Psi (V_{o })^T \Phi  Z_2 \label{adap2}\\
       &\dot{\hat{\mathcal{C}}}_{t }^{-1} =-\gamma_{3}  \Phi I_{t }Z_2\label{adap3}\\
   &\dot{\hat{L}}_{t_i }^{-1} =-\gamma_{4_i}  V_o(Z_2+\delta_iZ_{2i})\label{adap4}\\&\dot{\hat \lambda}_{i }=-\gamma_{5_i}  I_{t_i}(Z_2+\delta_iZ_{2i})\label{adap5}\\&\dot{\hat \mu}_{i }=\gamma_{6_i}  u_{i}(Z_2+\delta_iZ_{2i})\label{adap6}
\end{align}
 with  $\gamma_{2} $, $\gamma_{3} $, $\gamma_{4_i}$, $\gamma_{5_i}$ and $\gamma_{6_i} $ as positive scalars, where  $\delta_i=1$ for $i=1,...,n-1$ and $\delta_n=0$.    The following theorem summarizes the
design of the adaptive nonlinear controller:
\begin{thm}\label{theorem}
Consider the closed-loop system consisting of  the  dynamics  (\ref{z22}), (\ref{42ppiti}) and ( \ref{32ppiti1}),  the  control laws   (\ref{x2ref}), (\ref{nu_final}), and  (\ref{uii}) for $i=1,...,n-1$ and the adaptation laws   (\ref{adtet}), (\ref{adap2}), (\ref{adap3}),  (\ref{adap4}), (\ref{adap5}) and (\ref{adap6}) for $i=1,...,n$. 
 With   the gains    $\kappa_{1}$ and  $\kappa_{2}$, $\kappa_{2i}$  and the adaptation gains $\gamma_{1}$, $\gamma_{2}$, $\gamma_{3}$, $\gamma_{4_i}$, $\gamma_{5_i}$ and $\gamma_{6_i}$   for $i=1,...,n$  taking   arbitrary  positive values, if the voltage initial value satisfies  $ v_{min}<  V_{o }(0)< v_{max} $, then
 \begin{enumerate}[label=(\roman*)]
 \item All closed loop signals
are bounded for all $t\geq 0$. Moreover, we have   $ v_{min}<  V_{o }< v_{max} $  for all $t \geq 0$.
\item The output voltage $V_{o }$ and the current $I_{t_i}$ for $i=1,...,n$ converge to their setpoints asymptotically. 
\end{enumerate}
\end{thm}

\textit{\textbf{Proof:}}
Substituting  for $u$ and $u_i$ from (\ref{nu_final}) and (\ref{uii}) in (\ref{42ppiti}) and ( \ref{32ppiti1}), we  get the closed loop dynamics of $Z_2$ and $Z_{2i}$ as
\begin{align}
 \dot{Z}_{2}=&-\dfrac{d \mathcal{F}^{-1}(V_o)}{dV_o}Z_1-\kappa_2 Z_2+\Phi\Psi (V_{o })\tilde \Theta_{c }-\Phi I_{t }\tilde {\mathcal{C}}_{t }^{-1} \nonumber\\&-\Sigma_{i=1 }^{n}\tilde L_{t_{i }} ^{-1}V_{o}-\Sigma_{i=1 }^{n}\tilde \lambda_iI_{t_{i}}+\Sigma_{i=1 }^{n}\tilde\mu_iu_{i}\label{22ppiti}
 \\ \dot{Z}_{2i}=&-\kappa_{2i}Z_{2i}-  \tilde L_{t_{i }}^{-1} V_{o}- \tilde \lambda_iI_{t_{i}}+ \tilde \mu_iu_{i}\label{32ppiti}
  \end{align}
where $\tilde\Theta_{c}=\Theta_{c}-\hat\Theta_{c}$, $\tilde{\mathcal{C}}_{t}^{-1}=\mathcal{C}_{t}^{-1}-\hat{\mathcal{C}}_{t}^{-1}$,  $\tilde{L}_{t_i}^{-1}=L_{t_i}^{-1}-\hat{L}_{t_i}^{-1}$ and $\tilde{\lambda}_{ i}=\lambda_i  -\hat{\lambda}_{ i} $.  Now, consider  the following Lyapunov function candidate:
 \begin{align}
   & W  = W_1+\frac{1}{2}Z_2^2+\frac{1}{2}\Sigma_{i=1}^{n-1}  Z_{2i}^2 +\frac{1}{2}  \gamma_{2 }^{-1}\tilde \Theta_{c }^T \tilde \Theta_{c }+\frac{1}{2} \gamma_{3 }^{-1}(\tilde {\mathcal{C}}_{t }^{-1})^2 \nonumber\\&+\frac{1}{2}\Sigma_{i=1}^n \gamma_{4_i}^{-1}(\tilde L_{t_i}^{-1})^2+\frac{1}{2}\Sigma_{i=1}^n \gamma_{5_i}^{-1}\tilde \lambda_i^2+\frac{1}{2}\Sigma_{i=1}^n \gamma_{6_i}^{-1}\tilde \mu_i^2\label{2V2}
\end{align}
  where      $W_1$ is   in the form of (\ref{W1}). As shown in Section \ref{sec31}, considering the dynamic error  (\ref{z22}), if the virtual input $\xi$ is chosen as (\ref{x2ref}) and $\hat \Theta $ is calculated based on (\ref{adtet}), then $\dot W_1$ is upper bounded as (\ref{dddW1}). Differentiating $W$ with respect to time  and then substituting  for $\dot W_1$,  $\dot {Z}_{2}$ and  $\dot {Z}_{2i}$ from (\ref{dddW1}),   (\ref{22ppiti}) and ( \ref{32ppiti})  yield that 
 \begin{align}
   &  \dot W  = -\kappa_1 Z_{1}^2 -\kappa_2Z_2^2 -\Sigma_{i=1}^{n-1} \kappa_{2i}Z_{2i}^2\nonumber\\&+Z_2\Phi\Psi (V_{o })\tilde \Theta_{c }-Z_2\Phi I_{t }\tilde {\mathcal{C}}_{t }^{-1}\nonumber\\&-\Sigma_{i=1 }^{n}\tilde L_{t_{i }} ^{-1}V_{o}Z_2-\Sigma_{i=1 }^{n}\tilde \lambda_iI_{t_{i}}Z_2+\Sigma_{i=1 }^{n}\tilde\mu_iu_{i}Z_2 \nonumber\\&-\Sigma_{i=1}^{n-1}   \tilde L_{t_{i }}^{-1} V_{o}Z_{2i}- \Sigma_{i=1}^{n-1}\tilde \lambda_iI_{t_{i}}Z_{2i}+ \Sigma_{i=1}^{n-1}\tilde \mu_iu_{i} Z_{2i}\nonumber\\&-   \gamma_{2 }^{-1}\tilde \Theta_{c }^T \dot{\hat \Theta}_{c }- \gamma_{3 }^{-1} \tilde {\mathcal{C}}_{t }^{-1}\dot{\hat {\mathcal{C}}}_{t }^{-1}\nonumber\\&-\Sigma_{i=1}^n \gamma_{4_i}^{-1}\tilde L_{t_i}^{-1}\dot{\hat L}_{t_i}^{-1}-\Sigma_{i=1}^n \gamma_{5_i}^{-1}\tilde \lambda_i \dot{\hat \lambda}_i -\Sigma_{i=1}^n \gamma_{6_i}^{-1}\tilde \mu_i\dot{\hat \mu}_i  \label{Vv2}
\end{align}
Substituting  for $\dot {\hat \Theta}_{c}$,  $\dot{\hat {\mathcal{C}}}_{t }^{-1}$, $\dot{\hat {L}}_{t_i}^{-1}$, 
$\dot {\hat \lambda}_i$ and $\dot {\hat \mu}_i$ and doing  some simplifications, we get
\begin{align}
    \dot W  = -\kappa_1 Z_{1}^2 -\kappa_2Z_2^2 -\Sigma_{i=1}^{n-1} \kappa_{2i}Z_{2i}^2\label{2Vv2}
\end{align}

\begin{enumerate}[label=(\roman*)]

 \item Clearly, (\ref{2Vv2}) implies that    $\dot W \leq 0$. Therefore, it is readily concluded that $W(t)\leq W(0)$ for all $t>0$.  Since $ v_{min}<V_o(0)< v_{max}$, based on Remark \ref{rem1}, the initial value  $\mathcal{V}(0)$ and thus $W(0)$ are also bounded.  As a result, $W(t)$  and thus all the signals, including $\mathcal{V} $,  $I_t$, $I_{t_i}$, $ \hat \Theta_{c}$,   $ {\hat{\mathcal{C}}}_{t}^{-1}$,  ${\hat{L}}_{t_i}^{-1}$, $\hat \lambda_i$, $\hat \mu_i$ for $i=1,...,n$, remain bounded    for all $t\geq 0$. Consequently,  we know, from Remark \ref{rem1}, that $v_{min}< V_{o }(t)< v_{max}$ for all $t\geq 0$.    
\item  From the results in (i) , it is straightforward to show that $\dot {Z}_{1}$, $\dot {Z}_{2}$ and  $\dot {Z}_{2i}$, given in (\ref{z22}), (\ref{22ppiti}) and ( \ref{32ppiti}),  remain bounded.
Therefore, differentiating $\dot W $ in (\ref{2Vv2}),  we can show that $\ddot W $ is bounded, which means that $\dot W$
is uniformly continuous. Then, by Barbalat's Lemma \cite{slotine1991}, we conclude  that $Z_1$, $Z_2$ and $Z_{2i}$ for $i=1,...,n-1$  converge to zero asymptotically. Based on the error signals (\ref{itild})-(\ref{z2}), this implies that $V_{o } $, $I_t$ and $I_{t_i}$ converge  to $V_{o }^*$, $\xi$ and $\hat I_{t_i}^*=r_i\hat I_L^*$ for $i=1,...,n-1$. Based on (\ref{x2ref}), since $Z_1$ converges to zero, we conclude that $\xi$ converges to $\hat I_L^* = \Psi(V_o^*) \hat \Theta $. On the other hand, from (\ref{z22}), it is inferred that    $\xi$ converges to $  I_L^* = \Psi(V_o^*)  \Theta $. Therefore,    the estimated value $\hat I_L^*$ converges to  the actual current demand  $ I_L^*$. These results yield that $I_{t_i}$ converges  to $ r_i  I_L^*$ for $i=1,...,n-1$. Moreover, $I_{t_n}=I_L-\Sigma_{i=1}^{n-1} I_{t_i}$ converges to $I_{t_n}^*=I_{L}^*-\Sigma_{i=1}^{n-1} I_{t_i}^*= I_{L}^*-\Sigma_{i=1}^{n-1} r_i  I_L^*=(1-\Sigma_{i=1}^{n-1} r_i)I_L^*=r_nI_L^*$. Finally, the current sharing objective (\ref{ppiti}) is satisfied.\end{enumerate}
\begin{rmk} The virtual input $\xi$ in (\ref{x2ref}) is bounded if $\dfrac{d \mathcal{F}^{-1}(V_o)}{d V_o} \neq 0$. Therefore, barrier function $\mathcal{F}^{-1}(V_o)$ must be chosen such that  $\dfrac{d \mathcal{F}^{-1}(V_o)}{d V_o} \neq 0$ for $v_{min}< V_{o }(t)< v_{max}$. \end{rmk}

\begin{figure}
 \centering
  \includegraphics[width=6cm]{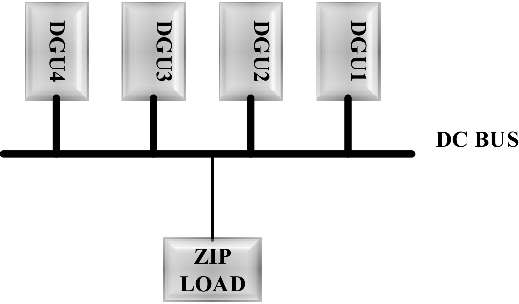} 
   \caption{Topology of the test DC system.} \label{topo}
 \end{figure}
\begin{table}
\begin{center}
\caption{Electrical parameters  of the test  system }
\begin{tabular} {|p{0.6cm}| p{0.8cm}| p{1cm}|p{1.2cm}|}
\hline
\multicolumn{4}{|c|}{Parameters of DG units} \\
\hline
 DGU&$E_i (V) $  & $R_{t_i}(\Omega) $ &  $L_{t_i}(mH) $ \\ \hline
1 & 24  & 0.1 & 1.3\\
\hline
2 & 24  & 0.1 & 1.2\\
\hline
3 & 24  & 0.1 & 1.6\\
\hline
4 & 24  & 0.1 & 1.4\\
\hline
\end{tabular}
\begin{tabular}{|p{1.28cm}| p{1.36cm}| p{1.36cm}|}
\hline
\multicolumn{3}{|c|}{Parameters of ZIP loads} \\
\hline
 $R_{l}(\Omega)$ & $I_{l}(A)$ & $P_{l} (W) $  \\ \hline
 1  & 5 & 120\\
\hline
\end{tabular}
\end{center}
\end{table}

 \section{Simulation Results}
In this section, we evaluate the performance of the proposed  controller via simulations performed in MATLAB/Simscape Electrical environment.  For this purpose, we consider a parallel converter system consisting of N = 4 DC-DC converters. The
system parameters are given in Table. II. Also, load capacitance is $\mathcal{C}_{t}=40 mF$, and the switching and sampling frequencies are $50kHz$ and $20KHz$, respectively. These parameters are
similar to those used in \cite{9201011}. 

We assume that the desired
  output voltage $V_o^*$ is $12 V$, with the safe region
between $v_{min}=11.8 V$ and $ v_{max}=12.2 V$. We consider the following  barrier function:
 \begin{align}
      V_o=\mathcal{F}(\mathcal{V})=\dfrac{v_{min}+v_{max}}{2}+\dfrac{v_{max}-v_{min}}{2}tanh(\mathcal{V}) \label{bbb1}
  \end{align}
  whose inverse is $\mathcal{V}=\mathcal{F}^{-1}(V_o)= \dfrac{1}{2}ln\Big(\dfrac{V_o-v_{min}}{v_{max}-V_o}\Big)$. Note that $\dfrac{d \mathcal{F}^{-1}(V_o)}{d V_o}=\dfrac{1}{2} \dfrac {v_{max}-v_{min}}{(v_{max}-V_o)(V_o-v_{min})} \neq 0 $ for  $ v_{min}<  V_{o }< v_{max} $.  We choose $r_1=0.4$, $r_2=0.3$, $r_3=0.2$ and $r_4=0.1$. We set the design parameter as $\kappa_{1}=1$, $\kappa_{2}=10$, $\kappa_{2i}=15$ and  $\gamma_{1 }=100$, $\gamma_{2}=100$, $\gamma_{3}=100$, $\gamma_{4_i }=100$, $\gamma_{5_i}=100$ and $\gamma_{6_i}=200$ for $i=1,...,4$.

\begin{figure}
 \centering
   \includegraphics[width=9.7 cm]{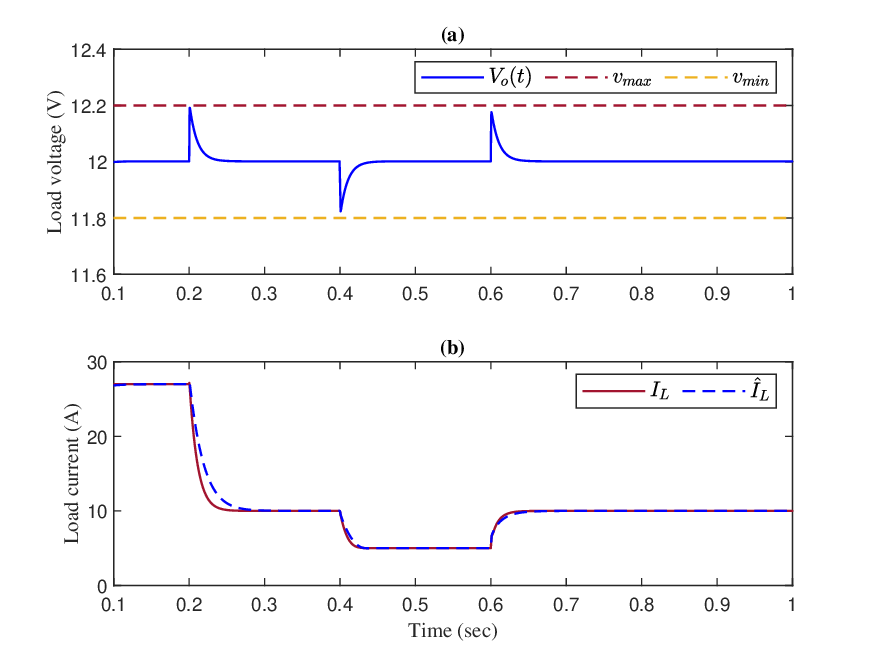}
   \caption{(a) Load voltage and (b) Load current. Z and I loads are disconnected   at $t=0.2~s$,   P-load changes at  $t=0.4~s$ from $120~W$ to $240~W$, and then changes at $t=0.6~s$ from $240~W$ to $120~W$.  } \label{fig_vref}
 \end{figure}

\begin{figure}
 \centering
   \includegraphics[width=9.7 cm]{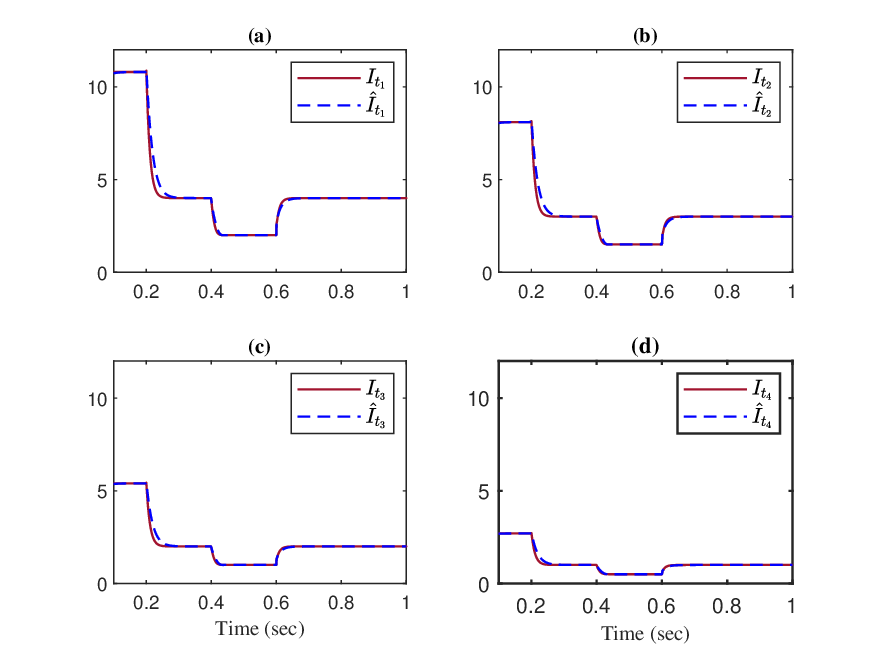}
   \caption{Output currents of the converters and their estimations.   Z and I loads are disconnected   at $t=0.2~s$,   P-load changes at  $t=0.4~s$ from $120~W$ to $240~W$, and then changes at $t=0.6~s$ from $240~W$ to $120~W$.  } \label{fig2_vref}
 \end{figure}

  We choose the initial value of the output voltage with the feasible voltage value. The proposed controller must be able to preserve its desired performance under large unknown variations of the loads.
 We examine the efficiency of the controllers under a large variation   in P-load
at a worst-case scenario. The worst-case scenario in terms of stability occurs when the Z-load is disconnected and a pure P-load is connected to the DC bus. Accordingly, we consider the case that the  Z-load and I-load  is disconnected at $t=0.2~s$ ( Z-load changes to a very large value
such as $1000000~\Omega$ and I-load changes to zero),  and thus DC microgrid feeds a pure  P-load with the value of $120~W$. We further step up this pure P-load from $120 ~W$ to $240 ~W$ at  $t=0.4~s$ and then we decrease it to $120 ~W$ at  $t=0.6~s$ to verify the effectiveness of our controller under  large variations of P-load at the worst-case scenario.  The output voltage  under these load variations is shown in Fig. \ref{fig_vref}~(a). As seen,  the output voltage undergo small variations at $t=0.2~s$, $t=0.4~s$, and  $t=0.6~s$  and is adjusted accurately to its setpoint value after a small transient. Also, output voltage does not exceed the consider safe range $[11.8,12.2] ~V$ despite these large load variations.    Moreover,  Fig. \ref{fig_vref} (b)  shows the load current estimation $\hat I_L$ converges to its actual value after the load change at $t=0.2~s$, $t=0.4~s$, and $t=0.6~s$. 
Fig. \ref{fig2_vref} shows the output currents of the converters and their estimations. These outcomes demonstrate that the closed-loop system remains stable even under pure P-load and   the proposed   controller is   robust to the large ZIP load variations. 
 \section{Conclusion}

In this paper,   we addressed the problem of voltage regulation and current sharing in the parallel DC-DC converter systems feeding common ZIP loads.   The proposed controller ensures simultaneous voltage adjustment and power sharing in the large signal sense despite uncertainties in ZIP loads, DC input voltages, and other electrical parameters. To keep the output voltage within a desired range, we utilized a barrier  function. We applied  the invertible transformation of the barrier function on the output voltage and then designed  the  controller using the adaptive backstepping method.
The effectiveness of the
results was illustrated via simulations of a parallel DC-DC converter system.

\bibliographystyle{IEEEtran}
\bibliography{library}

\begin{thebibliography}{10}
\providecommand{\url}[1]{#1}
\def\UrlFont{\rmfamily}
\providecommand{\newblock}{\relax}
\providecommand{\bibinfo}[2]{#2}
\providecommand\BIBentrySTDinterwordspacing{\spaceskip=0pt\relax}
\providecommand\BIBentryALTinterwordstretchfactor{4}
\providecommand\BIBentryALTinterwordspacing{\spaceskip=\fontdimen2\font plus
\BIBentryALTinterwordstretchfactor\fontdimen3\font minus \fontdimen4\font\relax}
\providecommand\BIBforeignlanguage[2]{{%
\expandafter\ifx\csname l@#1\endcsname\relax
\typeout{** WARNING: IEEEtran.bst: No hyphenation pattern has been}%
\typeout{** loaded for the language `#1'. Using the pattern for}%
\typeout{** the default language instead.}%
\else
\language=\csname l@#1\endcsname
\fi
#2}}

\bibitem{8616804}
F.~Chen, R.~Burgos, D.~Boroyevich, J.~C. Vasquez, and J.~M. Guerrero, ``Investigation of nonlinear droop control in dc power distribution systems: Load sharing, voltage regulation, efficiency, and stability,'' \emph{IEEE Transactions on Power Electronics}, vol.~34, no.~10, pp. 9404--9421, 2019.

\bibitem{Anand}
S.~Anand and B.~G. Fernandes, ``Modified droop controller for paralleling of dc–dc converters in standalone dc system,'' \emph{IET Power Electronics}, vol.~5, no.~6, pp. 782--789, 2012.

\bibitem{988666}
J.-W. Kim, H.-S. Choi, and B.~H. Cho, ``A novel droop method for converter parallel operation,'' \emph{IEEE Transactions on Power Electronics}, vol.~17, no.~1, pp. 25--32, 2002.

\bibitem{704129}
V.~Thottuvelil and G.~Verghese, ``Analysis and control design of paralleled dc/dc converters with current sharing,'' \emph{IEEE Transactions on Power Electronics}, vol.~13, no.~4, pp. 635--644, 1998.

\bibitem{6851919}
S.~Moayedi, V.~Nasirian, F.~L. Lewis, and A.~Davoudi, ``Team-oriented load sharing in parallel dc–dc converters,'' \emph{IEEE Transactions on Industry Applications}, vol.~51, no.~1, pp. 479--490, 2015.

\bibitem{4195635}
Y.~Huang and C.~K. Tse, ``Circuit theoretic classification of parallel connected dc–dc converters,'' \emph{IEEE Transactions on Circuits and Systems I: Regular Papers}, vol.~54, no.~5, pp. 1099--1108, 2007.

\bibitem{6727450}
H.~Behjati, A.~Davoudi, and F.~Lewis, ``Modular dc–dc converters on graphs: Cooperative control,'' \emph{IEEE Transactions on Power Electronics}, vol.~29, no.~12, pp. 6725--6741, 2014.

\bibitem{4418525}
S.~K. Mazumder, M.~Tahir, and K.~Acharya, ``Master–slave current-sharing control of a parallel dc–dc converter system over an rf communication interface,'' \emph{IEEE Transactions on Industrial Electronics}, vol.~55, no.~1, pp. 59--66, 2008.

\bibitem{6109354}
J.~Shi, L.~Zhou, and X.~He, ``Common-duty-ratio control of input-parallel output-parallel (ipop) connected dc–dc converter modules with automatic sharing of currents,'' \emph{IEEE Transactions on Power Electronics}, vol.~27, no.~7, pp. 3277--3291, 2012.

\bibitem{5457984}
D.~Sha, Z.~Guo, and X.~Liao, ``Cross-feedback output-current-sharing control for input-series-output-parallel modular dc–dc converters,'' \emph{IEEE Transactions on Power Electronics}, vol.~25, no.~11, pp. 2762--2771, 2010.

\bibitem{4745797}
P.~J. Grbovic, ``Master/slave control of input-series- and output-parallel-connected converters: Concept for low-cost high-voltage auxiliary power supplies,'' \emph{IEEE Transactions on Power Electronics}, vol.~24, no.~2, pp. 316--328, 2009.

\bibitem{5613924}
M.~Li, C.~K. Tse, H.~H.~C. Iu, and X.~Ma, ``Unified equivalent modeling for stability analysis of parallel-connected dc/dc converters,'' \emph{IEEE Transactions on Circuits and Systems II: Express Briefs}, vol.~57, no.~11, pp. 898--902, 2010.

\bibitem{8269340}
R.~Delpoux, J.-F. Trégouët, J.-Y. Gauthier, and C.~Lacombe, ``New framework for optimal current sharing of nonidentical parallel buck converters,'' \emph{IEEE Transactions on Control Systems Technology}, vol.~27, no.~3, pp. 1237--1243, 2019.

\bibitem{TREGOUET201959}
J.-F. Trégouët and R.~Delpoux, ``New framework for parallel interconnection of buck converters: Application to optimal current-sharing with constraints and unknown load,'' \emph{Control Engineering Practice}, vol.~87, pp. 59--75, 2019.

\bibitem{8960535}
J.~Kreiss, J.-F. Trégouët, D.~Eberard, R.~Delpoux, J.-Y. Gauthier, and X.~Lin-Shi, ``Hamiltonian point of view on parallel interconnection of buck converters,'' \emph{IEEE Transactions on Control Systems Technology}, vol.~29, no.~1, pp. 43--52, 2021.

\bibitem{8064725}
M.~A. Setiawan, A.~Abu-Siada, and F.~Shahnia, ``A new technique for simultaneous load current sharing and voltage regulation in dc microgrids,'' \emph{IEEE Transactions on Industrial Informatics}, vol.~14, no.~4, pp. 1403--1414, 2018.

\bibitem{Singh2017}
S.~Singh, A.~R.Gautam, and D.~Fulwani, ``Constant power loads and their effects in {DC} distributed power systems: A review,'' \emph{Renewable and Sustainable Energy Reviews}, vol.~72, pp. 407--421, 2017.

\bibitem{9201011}
M.~S. Sadabadi, ``A distributed control strategy for parallel dc-dc converters,'' \emph{IEEE Control Systems Letters}, vol.~5, no.~4, pp. 1231--1236, 2021.

\bibitem{9611083}
P.~Nahata, M.~S. Turan, and G.~Ferrari-Trecate, ``Consensus-based current sharing and voltage balancing in dc microgrids with exponential loads,'' \emph{IEEE Transactions on Control Systems Technology}, vol.~30, no.~4, pp. 1668--1680, 2022.

\bibitem{10471262}
A.~J. Malan, P.~Jané-Soneira, F.~Strehle, and S.~Hohmann, ``Passivity-based power sharing and voltage regulation in dc microgrids with unactuated buses,'' \emph{IEEE Transactions on Control Systems Technology}, pp. 1--16, 2024.

\bibitem{8836491}
B.~Fan, S.~Guo, J.~Peng, Q.~Yang, W.~Liu, and L.~Liu, ``A consensus-based algorithm for power sharing and voltage regulation in dc microgrids,'' \emph{IEEE Transactions on Industrial Informatics}, vol.~16, no.~6, pp. 3987--3996, 2020.

\bibitem{9279289}
J.~Peng, B.~Fan, Q.~Yang, and W.~Liu, ``Fully distributed discrete-time control of dc microgrids with zip loads,'' \emph{IEEE Systems Journal}, vol.~16, no.~1, pp. 155--165, 2022.

\bibitem{DEPERSIS2018364}
C.~DePersis, E.~R. Weitenberg, and F.~Dörfler, ``A power consensus algorithm for dc microgrids,'' \emph{Automatica}, vol.~89, pp. 364--375, 2018.

\bibitem{9409144}
K.~C. Kosaraju, S.~Sivaranjani, and V.~Gupta, ``Safety during transient response in direct current microgrids using control barrier functions,'' \emph{IEEE Control Systems Letters}, vol.~6, pp. 337--342, 2022.

\bibitem{9991257}
S.~Singh, V.~Vaishnav, A.~Jain, and D.~Sharma, ``Bounded voltage regulation in a direct current microgrid using barrier lyapunov function with uncertain load current,'' \emph{IEEE Control Systems Letters}, vol.~7, pp. 991--996, 2023.

\bibitem{10412654}
S.~Bahrami, S.~S. Mousavi, and M.~Shafiee-Rad, ``Decentralized adaptive nonlinear controller for voltage regulation of output-constrained dc microgrids with zip loads,'' \emph{IEEE Transactions on Power Systems}, pp. 1--12, 2024.

\bibitem{Soloperto2018}
R.~Soloperto, P.~Nahata, M.~Tucci, and G.Ferrari-Trecate, ``A passivity-based approach to voltage stabilization in {DC} microgrids,'' in \emph{American control conference(ACC)}, 2018, pp. 5374--5379.

\bibitem{24nahata2020}
P.~Nahata, R.~Soloperto, M.~Tucci, A.~Martinelli, and G.~Ferrari-Trecate, ``A passivity-based approach to voltage stabilization in {DC} microgrids with {ZIP} loads,'' \emph{Automatica}, vol. 113, no.~6, p. 108770, 2020.

\bibitem{9134402}
J.~Ferguson, M.~Cucuzzella, and J.~M.~A. Scherpen, ``Exponential stability and local {ISS} for {DC} networks,'' \emph{IEEE Control Systems Letters}, vol.~5, no.~3, pp. 893--898, 2021.

\bibitem{slotine1991}
J.~E. Slotine and W.~Li, \emph{Applied nonlinear control}.\hskip 1em plus 0.5em minus 0.4em\relax Englewood Cliff, NJ: Prentice-Hall, 1991.

\end{thebibliography}

\end{document}